# Influence of Auger heating and Shockley-Read-Hall recombination on hot carrier dynamics in InGaAs nanowires

Hamidreza Esmaielpour[*], Nabi Isaev, Jonathan J. Finley, Gregor Koblmüller[*]

Walter Schottky Institut, TUM School of Natural Sciences, Technical University of Munich, 85748 Garching, Germany.

**Abstract** – Understanding the origin of hot carrier relaxation in nanowires (NWs) with one-dimensional (1D) geometry is significant for designing efficient hot carrier solar cells with such nanostructures. Here, we study the influence of Auger heating and Shockley-Read-Hall recombination on hot carrier dynamics of catalyst-free InGaAs-InAlAs core-shell NWs. Using steady-state and time-resolved photoluminescence (PL) spectroscopy the dependencies of hot carrier effects on the degree of confinement of photo-generated carriers induced by the nanowire diameter are determined at different lattice temperatures. Analysis of excitation-power dependent data and temperature-dependent PL linewidth broadening reveal that at low temperatures, strong Auger recombination and phonon-bottleneck are responsible for hot carrier effects. Our analysis gives also insights into electron-phonon and ionized impurity scattering, showing opposing trends with NW diameter, and it allows to estimate the Fröhlich coupling constant for the InGaAs NWs. Conversely, with increasing lattice temperature, hot carrier relaxation rates increase due to enhanced Shockley-Read Hall and surface recombination. Time-resolved spectroscopy reveals a fourfold increase in the rate of Shockley-Read-Hall recombination from 6 ns at 10 K to 1.5 ns at 150 K. The findings suggest that minimizing defect densities in the bulk and surfaces of these NWs will be key to enhance hot carrier effects towards higher temperatures.

## I. INTRODUCTION

To boost the efficiency of photovoltaic solar cells beyond their theoretical limit in single-junction solar cells, it requires to suppress one of the main energy loss mechanisms in these devices, called thermalization of hot carriers. Understanding the origin of hot carrier relaxation in solar cells and inhibiting thermalization pathways is essential to fabricate efficient hot carrier solar cells.

One of the main relaxation pathways of hot carriers in polar (group III-V) semiconductors is via electron interactions with longitudinal optical (LO) phonons, known as Fröhlich interactions [1]. This process occurs quickly after photo-generation ($\sim ps$), however, it is possible to inhibit the dissipation of excess kinetic energy of hot carriers via creating a phonon-bottleneck effect [2,3]. Through this mechanism non-equilibrium distributions of phonons, i.e., hot phonons, are generated, which can slow down the rates of hot carrier relaxation by exchanging energy with these particles [4,5].

Nanostructures have been reported as one of the promising materials for the generation of hot carriers via phonon-bottleneck effects [6,7,8,9]. Nanowires (NWs), in particular, are a class of nanostructures that have shown evidence of strong hot carrier effects, given their spatial confinement in two dimensions and creation of even more significant phonon-bottleneck effects [10,11,12]. In addition, due to their large surface-to-volume ratio, they can enhance internal light reflection in their structures, which can increase photo-absorption and the density of hot carriers in the system [13,14]. Evidence of hot carrier effects in NWs was studied by steady-state photoluminescence (PL) [15,16], time-resolved pump-probe spectroscopy [17,18,19], and electrical measurements [20,21,22]. It has been shown that by reducing the diameter of the NWs and increasing the confinement of carriers in the system, the temperature of hot carriers increases due to the suppression of energy dissipation via phonon-mediated channels [15,16,23].

---

[*] Email address: hamidreza.esmaielpour@wsi.tum.de
Email address: gregor.koblmueller@wsi.tum.de

Another mechanism, that can lead to hot carrier generation is Auger recombination, or Auger heating [18,24,25]. This effect can be even more pronounced in nanostructures due to the low dimensionality of these structures [26]. By increasing the excitation power, the probability of Auger recombination increases due to the large excess carrier densities generated. Although this recombination mechanism is non-radiative and reduces the density of carriers, it is considered as one source of hot carrier generation, which can improve the efficiency of hot carrier photovoltaic solar cells [27].

Despite significant advantages of nanostructures for the generation of hot carriers, the proximity of interfaces and surfaces can affect the relaxation rates of hot carriers by surface recombination, strain, and alloy fluctuations at the interface [28,29]. It has been shown that by passivating dangling bonds on the surface using lattice-matched compounds, it is possible to reduce the rates of hot carrier relaxation in the system and improve the yield of luminescence in these structures [18,30,31]. To what extent non-radiative processes, such as Auger heating and surface recombination, influence the hot carrier effects in NWs, is however largely unknown.

Here, we study the properties of hot carriers in core-shell InGaAs/InAlAs NWs via steady-state and time-resolved micro-photoluminescence (µPL) spectroscopy to reveal the impact of non-radiative recombination processes. We first show from steady-state PL experiments performed at various lattice temperatures how the hot carrier properties are described by the temperature-dependent PL linewidth broadening and the associated scattering mechanisms for a variety of different NW diameters. The results are further correlated by a comprehensive study of carrier recombination dynamics via time-resolved PL spectroscopy to elucidate the influence of Auger and Shockley-Read-Hall (SRH) recombination mechanisms on the dynamics of hot carriers.

## II.     EXPERIMENTAL METHODS AND DISCUSSIONS

The NWs studied in this work consist of closely lattice-matched $In_{0.2}Ga_{0.8}As/In_{0.2}Al_{0.8}As$ core-shell NWs, that were grown by molecular beam epitaxy (MBE) using a catalyst-free, selective-area epitaxial growth method. Different sets of NW samples were explored with varying InGaAs core diameter (110-200 nm), under otherwise fixed InAlAs shell thickness (~10 nm). Details of their growth is presented elsewhere [16,30,32]. Figure 1(a) shows the PL spectra emitted by the NWs (160 nm InGaAs core diameter in this case) at various lattice temperatures. A characteristic redshift behavior is observed by increasing temperature, as expected due to thermally induced lattice expansion and reduction of the band-gap energy. As a measure for the hot carrier effect, the temperature of photo-generated hot carriers is determined by fitting the PL spectra using the absorptivity method [26] with the generalized Planck's radiation law, as given by [33,34]:

$$I_{PL}(E) = \frac{2\pi A(E) (E)^2}{h^3 c^2} \left[ exp\left(\frac{E - \Delta\mu}{k_B T}\right) - 1 \right]^{-1}. \qquad (1)$$

Here, "$I_{PL}$" is the PL intensity, "$A(E)$" the energy-dependent absorptivity, "$h$" the Planck constant, "$k_B$" the Boltzmann constant, "$c$" the speed of light. "$T$" and "$\Delta\mu$" are the carrier temperature and the quasi-Fermi level splitting, respectively. The results of the hot carrier temperature (ΔT: temperature difference between the hot carrier and the lattice temperature) versus the lattice temperature at 6 kW/cm² absorbed power density are shown in Figure 1(b) and compared with various other NW-diameters. It is seen that at 10 K, the maximum of hot carrier temperature is observed in the NWs of 160 nm diameter. In addition, it is seen that by increasing the lattice temperature, the effect of hot carriers reduces. This behavior is observed in all NWs; however, the thinner NWs lose their hot carrier effects faster than the thicker NWs. To identify the origins for this intriguing behavior, and especially the different dissipation strengths of the hot carrier effect with diameter, a closer investigation of the dominant recombination mechanism is performed.

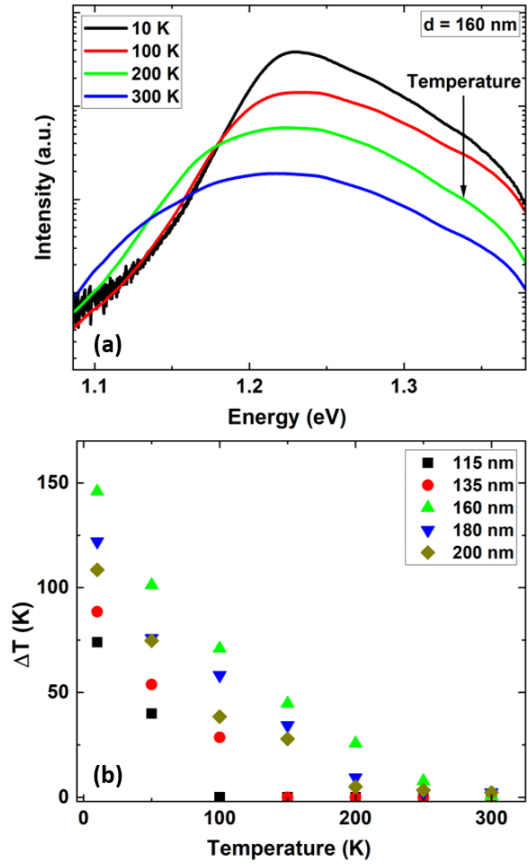

Figure 1. (a) Temperature-dependent PL spectra of the InGaAs/InAlAs NWs of 160 nm diameter recorded at 6 kW/cm² absorbed power density. (b) Corresponding ΔT versus the lattice temperature for NWs of various diameters.

To study the dynamics of hot carriers under steady-state condition, we therefore explore the recombination mechanisms using the rate equation, as given by [35]:

$$P_{Abs.} = A\, I_{PL}^{1/2} + B\, I_{PL} + C\, I_{PL}^{3/2}, \qquad (2)$$

where "$A$", "$B$", and "$C$" are the coefficients attributed to Shockley-Read-Hall (SRH), radiative, and Auger recombination mechanisms, respectively. The dominant recombination mechanism in the system can be determined by finding the slope of the natural logarithm of the absorbed power density versus the natural logarithm of the integrated PL intensity [26]. Such analysis is performed in Figure 2(a) for the InGaAs NWs of 115 nm, 135 nm, and 180 nm diameter. It is seen that at 10 K under high excitation power density, where hot carrier effects are most evident, Auger recombination has the highest contribution in carrier recombination in the NWs given the large slope of ~1.5. The associated plots at higher lattice temperatures are shown in the supplementary information.

Figure 2(b) presents the dependence of the slopes extracted at high absorbed power densities, see the highlighted region, versus the lattice temperature. It is seen that by increasing the lattice temperature, the magnitude of the slope decreases rapidly, and at elevated temperatures, it shifts towards dominant SRH recombination (slope approximating ~ 0.5). The results indicate that by increasing the lattice temperature, the contribution of Auger heating reduces, leading to weaker hot carrier effects in the NWs, in agreement with Figure 1(b).

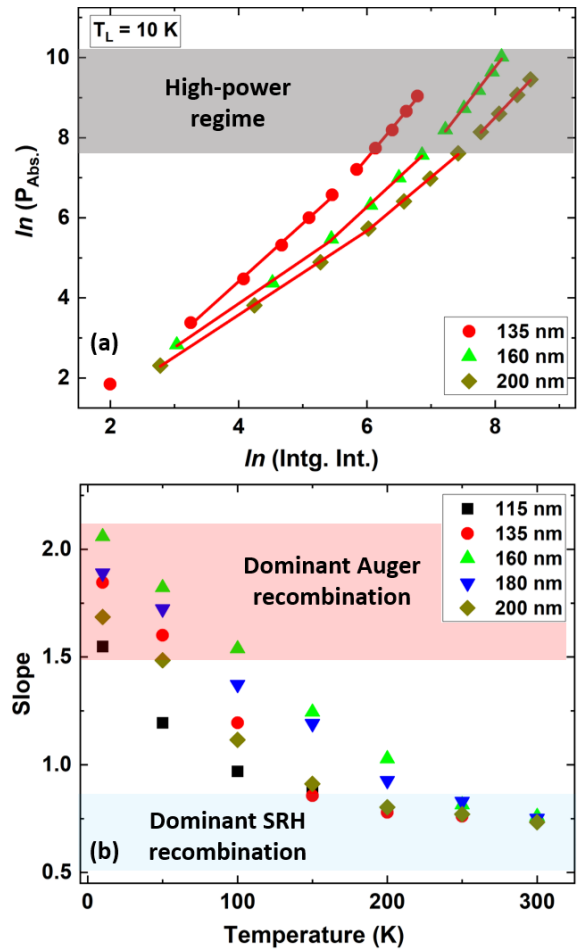

Figure 2. (a) Natural logarithm of absorbed power density versus the natural logarithm of integrated PL intensity at 10 K for NWs of 115 nm, 135 nm and 160 nm diameter. The dashed lines indicate the slopes of the data. (b) Slopes of the power-law analysis indicated in panel (a) at high excitation powers (highlighted region).

The increased contribution from SRH-recombination at elevated temperature motivates to take a closer look at the scattering dynamics of hot carriers and to identify the underlying defect-mediated and electron-phonon coupling effects. Such information about the scattering mechanism in the system can be assessed by analysis of the temperature-dependent PL linewidth broadening, Γ(T), as given by [36]:

$$\Gamma(T) = \Gamma_0 + \Gamma_{LA}T + \frac{\Gamma_{LO}}{[exp(h\omega_{LO}/k_BT) - 1]} + \Gamma_{imp}exp\left[-\frac{E_B}{k_BT}\right], \qquad (3)$$

where "$\Gamma_0$" is the inhomogeneous (defect) broadening, "$\Gamma_{LA}$" the coefficient due to longitudinal acoustic scattering, "$\Gamma_{LO}$" the Fröhlich coupling constant, "$h\omega_{LO}$" the energy of longitudinal optical (LO) phonons, "$T$" the lattice temperature, "$\Gamma_{imp}$" the ionized impurity coefficient, and "$E_B$" the impurity binding energy. The contribution of each scattering mechanism can be analyzed by fitting the experimental results with Equation 3.

The results of temperature-dependent PL linewidth broadening are plotted in Figure 3(a). The magnitude of the spectral linewidth is taken from the half-width-at-half-maximum (HWHF) of the low-energy side of the PL spectra emitted under low-excitation powers in order to exclude the contribution of hot carriers on the linewidth broadening. By fitting the experimental results with equation 3, it is possible to determine the contribution of various scattering mechanisms in the InGaAs NWs. The solid lines in Figure 3(a) indicate the best fits to the data of the temperature-dependent PL linewidth broadening for the NWs of various diameters. It is seen that by increasing the NW diameter, the magnitude of the inhomogeneous broadening reduces. The origin of this effect is attributed to a reduction of the defect density in the NWs grown by the catalyst-free growth technique at larger diameters [37,38].

Figure 3(b) shows the results of the ionized impurity coefficient and the Fröhlich coupling constant of the NWs determined by the linewidth fitting versus their diameter. In this analysis, the coefficient of longitudinal acoustic phonon scattering is found to be 0.02 meV/K, the LO-phonon energy, 36 meV, and the binding energy, 8.2 meV, for the InGaAs NWs with various diameters, which agree with reported values for similar compounds in literature [39,40]. It is observed that by increasing the diameter of the NWs, the coefficient of carrier scattering via ionized impurities reduces, while the Fröhlich coupling constant increases. The larger contribution of the ionized impurities for the thinner NWs is attributed to their higher density of defects [16], which are expected to introduce larger amounts of residual impurities due to segregation effects [41,42]. By increasing the lattice temperature, the impurities in these trap states will be ionized via receiving thermal energy and interact with photogenerated carriers, resulting in larger linewidth broadening of the PL spectrum versus temperature [43,44]. However, by increasing the diameter of the NWs and improving the crystal quality, the electron-phonon coupling increases via Fröhlich interactions. This effect is important for hot carrier absorbers, as it leads to higher probability of energy exchange between electrons (or holes) with LO-phonons, which can provide the conditions for the creation of the phonon-bottleneck effect in the system [2,45]. These results agree with the ab initio simulations on electron-phonon coupling in CdSe quantum dots, which showed that the introduction of defects into this structure can alter the electron-phonon scattering rates, leading to weaker phonon-bottleneck effects in the system [28]. In addition, the dependence of hot carrier relaxation rates on the density of defects was also studied in AlGaN/GaN heterostructures by Hall measurements and it was found that strong scattering of electrons with impurities and defects accelerates carrier relaxation rates, resulting in weaker hot carrier effects [29]. Theoretical and experimental results indicate that to design an efficient hot carrier absorber with robust phonon-bottleneck effects, it is necessary to improve the crystal quality and to reduce the density of structural defects in the system.

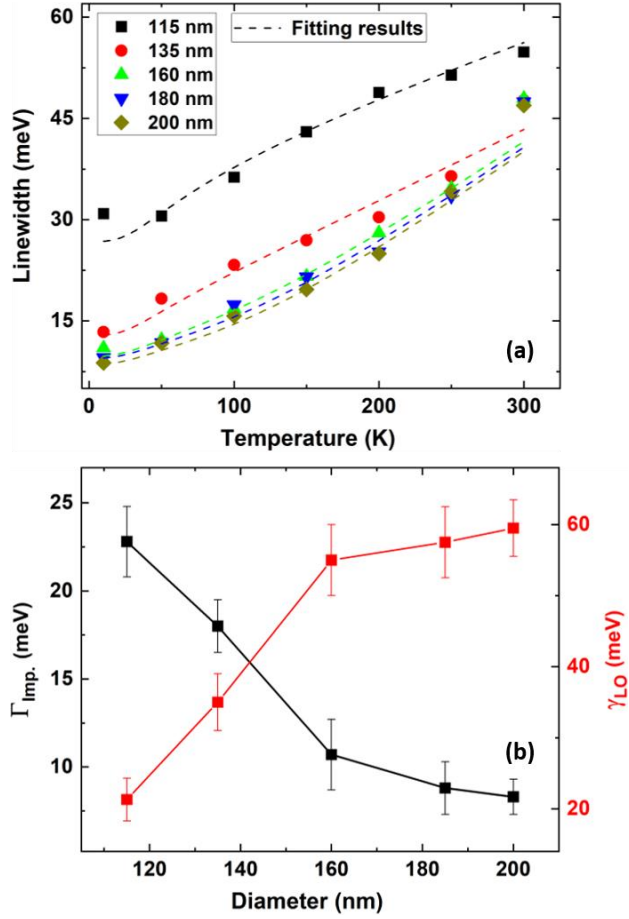

Figure 3. (a) Temperature-dependent PL linewidth broadening for the InGaAs NWs of different diameters. The dashed lines indicate the fitting results of the PL linewidth broadening versus the lattice temperature. (b) Dependence of ionized impurity scattering and electron-LO phonon coupling constant versus the NW diameter.

So far, steady-state PL spectroscopy has shown evidence of strongest hot carrier effects in the NWs of 160 nm. The origin of this effect is attributed to the competition between the phonon-bottleneck and Auger heating, on the one hand – resulting stronger hot carrier effects by reducing the NW diameter – and on the other hand, the increase of hot carrier relaxation rates due to higher density of defects in the thin NWs. Therefore, it is expected that by reducing the NW diameter, the rate of carrier recombination via the SRH mechanism increases. To underpin these expectations, we performed time-resolved PL (TRPL) spectroscopy at various lattice temperatures and excitation powers on the InGaAs NWs with various diameters. To determine the contributions of various recombination mechanisms, we apply the common rate equation for time-dependent carrier concentrations (n(t) and p(t) for electrons and holes, respectively), which is given by [46]:

$$\frac{d\Delta n}{dt} = -C_n\, n(np - n_i^2) - C_p\, p(np - n_i^2) - k_{rad}\, p(np - n_i^2) - \frac{np - n_i^2}{\tau_p n + \tau_n p}, \qquad (4)$$

where "$C_n$" and "$C_p$" are the Auger coefficients for electrons and holes, "$k_{rad}$" is the radiative recombination coefficient, "$n_i$" the intrinsic carrier density, and "$\tau_p$" and "$\tau_n$" are the non-radiative SRH recombination lifetimes, respectively. Under the approximation of high-excitation powers and considering $n = p$, Equation 4 can be written as [47],

$$\frac{d\Delta n}{dt} = -(C_n + C_p)n^3 - k_{rad}\,n^2 - \frac{n}{(\tau_p + \tau_n)}, \qquad (5)$$

where the first term is due to the Auger recombination, the second one due to the radiative recombination and the third one due to the SRH recombination mechanism. In this relationship, under the absence of Auger and radiative recombination, a mono-exponential decay of the TRPL signal occurs under relatively low excitation power densities and the signal follows a linear dependence between the natural logarithm of the TRPL intensity and the delay time [46]. Figure 4(a) examines these expectations by recording excitation power-dependent TRPL spectra at 10 K for InGaAs NWs (diameter of 160 nm), using a 780 nm Ti-Sapphire pulsed laser excitation with an 80 MHz repetition rate. The TRPL signal is recorded by a time-correlated single photon counting (TCSPC) system, which consists of a single-photon avalanche diode (SPAD) connected to a PicoHarp300 detector. The instrument response function (IRF) of the system is shown by a dashed line in Figure 4(a). It is seen that by increasing the absorbed power density, the slope of the TRPL spectrum at earlier decay times becomes steeper.

To quantitatively assess the contribution of various recombination mechanisms and Auger heating on hot carrier dynamics, one can determine the time constant of the TRPL signal by taking the derivative of the natural logarithm of the intensity ($\phi_{TRPL}$) versus the delay time, as given by [46]:

$$\tau_{TRPL} = \left(-\frac{d\,ln(\phi_{TRPL})}{dt}\right)^{-1}. \qquad (6)$$

According to Equation 6, under a mono-exponential decay, the lifetime determined by this equation represents the SRH recombination lifetime. Under high excitation powers, the contributions of Auger and radiative recombination increases, and the decay profile of the TRPL signal will not be mono-exponential; in other words, the spectrum cannot be described by only one time-constant. Therefore, by determining the power-dependent carrier lifetime, it is possible to study the impact of Auger and radiative processes on the carrier dynamics. It is expected that in a system with stronger Auger recombination, the slope of the TRPL decay becomes steeper, leading to shorter carrier lifetimes [48]. Since the rate of carrier recombination due the Auger mechanism has a cubic dependence on the carrier concentration, one should determine the lifetime right after the decay starts. Therefore, the dependence of lifetime on the excitation power is determined after 650 ps, as shown by the shaded area in Figure 4(a), which is larger than the FWHM of the IRF profile, to exclude its contribution from the TRPL signal.

The results of the power-dependent decay rate (inverse of the lifetime) versus the excitation power density for the InGaAs NWs of various diameters are plotted in Figure 4(b) for the case of low temperature (10 K). We observe that the decay rate of the 160 nm thick NWs has the largest values as a function of excitation power density. Recalling Figure 1(b), it was found that the InGaAs NWs of 160 nm diameter show the strongest hot carrier effects compared with other NWs, and the contribution of Auger heating determined from steady-state PL spectroscopy was dominant in these NWs (see Figure 2). To compare the results of steady-state and time-resolved PL data, the decay rate of carriers under 5 kW/cm² absorbed power density and the slope of the rate equation found by steady-state PL experiments for the Auger recombination versus the NW diameter are plotted in Figure 4(c). Both results exhibit the same trends, i.e., largest decay rates and strongest Auger heating effect for intermediate NW-diameter of 160 nm. This good agreement confirms the contribution of dominant Auger recombination on hot carrier dynamics at intermediate NW diameters in the limit of low lattice temperature.

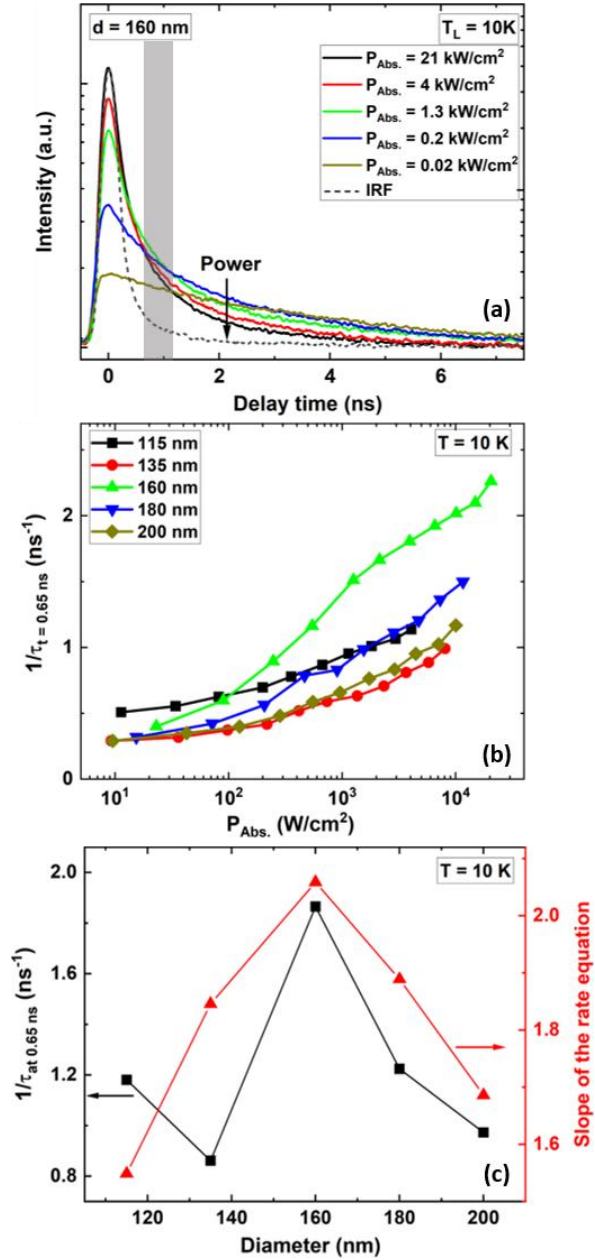

Figure 4. (a) TRPL spectra for the InGaAs NWs with 160 nm diameter at 10 K under various excitation powers. The shaded area indicates the region, where the contribution of Auger decay rate is determined. (b) Decay rate (inverse of lifetime) of carriers versus the excitation power at 10 K for the InGaAs NWs with various diameters. (c) Dependence of carrier decay rates at 5 kW/cm$^2$ extracted from the TRPL data and the slope of the rate equation determined by steady-state PL experiments versus the NW diameter.

The situation changes when increasing the lattice temperature, as the contribution of SRH recombination increases (cf. Figure 2(b)), resulting in higher rates of hot carrier relaxation. To determine the dependence of carrier lifetime due to the SRH recombination versus the lattice temperature, the TRPL spectra under low-excitation powers are analyzed, where the contribution of the SRH recombination is dominant [46]. The results of the SRH recombination lifetime versus the lattice temperature are plotted in Figure 5(a). It is seen that by increasing the lattice temperature, the recombination lifetime reduces from 6.5 ns at 10 K to 1.5 ns at 150 K for the NWs with 160 nm, indicative of higher rates of the SRH recombination at elevated temperatures. Similar behavior is observed for other InGaAs NWs with different diameters. At elevated lattice temperatures (> 150 K), the SRH recombination lifetime for the NWs could not be determined due to the IRF of the TRPL setup. These results agree with the drop of the integrated steady-state PL intensity versus the lattice temperature, see Figure S2 in the supporting information. In addition, Figure 5(a) indicates that the NWs with larger diameters, whose density of defects is less, have longer carrier lifetimes, indicative of their lower SRH recombination rates compared to the thinner NWs.

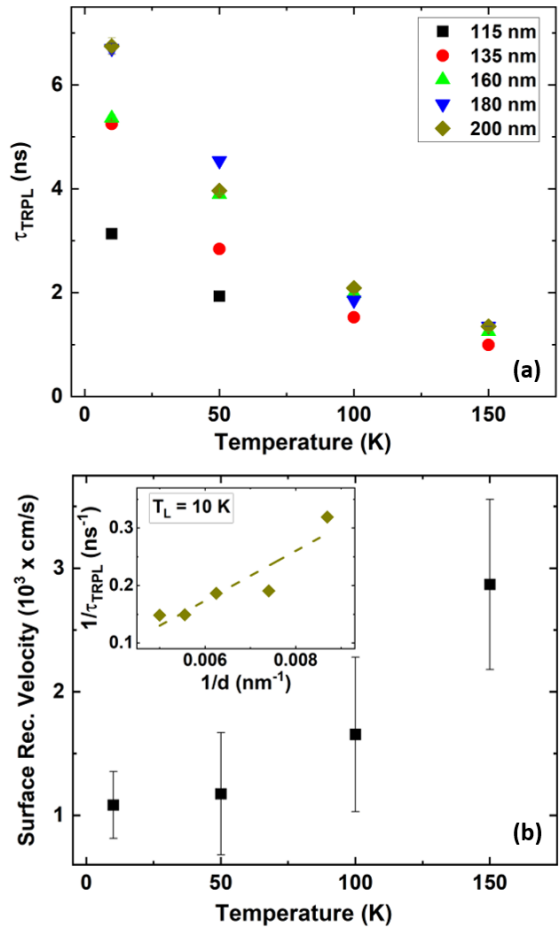

Figure 5. (a) Carrier lifetime of the InGaAs NWs for various diameters under low-absorbed power densities (10 W/cm$^2$) versus the lattice temperature. (b) Surface recombination velocity of the InGaAs NWs versus the lattice temperature. The inset of the graph shows the inverse of the TRPL lifetime under low-excitation power density, taken from panel (a), versus the inverse of the NW diameter at 10 K. The dashed line is a best fit to the data.

In nanostructures with small dimensions, surface recombination at the heterointerfaces or with dangling bonds at the surface can have large contributions to the carrier recombination rates [49]. It is expected that by reducing the dimensions of the NWs, the rates of carrier recombination at the surface increases. Concurrently, a system with large surface recombination velocities can increase the rates of hot carrier relaxation, leading to weaker hot carrier effects [50,51]. The carrier lifetime under low-excitation powers, $\tau_{TRPL}$, consists of a combination of the carrier lifetime in the bulk, $\tau_{Bulk}$, and the surface recombination velocity. To determine the contribution of the surface recombination velocity ($v_s$), one can fit the experimental results by [52]:

$$\tau_{TRPL}^{-1} = \left(\frac{4}{d}\right) v_s + \tau_{Bulk}^{-1}. \qquad (7)$$

In this relationship, by plotting the inverse of TRPL time constant under low-excitation powers versus the inverse of the NW diameter, the slope indicates the surface recombination velocity, as shown by the inset of Figure 5(b). The resulting surface recombination velocity in the InGaAs NWs, that is independent of the NW diameter, versus the lattice temperature is plotted in Figure 5(b). It is observed that by increasing the lattice temperature, the surface recombination velocity increases from $1\times10^3$ cm/s at 10 K to $\sim 3\times10^3$ cm/s at 150 K. These values agree with reported literature data for passivated GaAs/AlGaAs NWs [53] and about seven times larger than the ultralow surface recombination velocity reported for InGaAs/InP NWs [54] and InP NWs [55]. The increased surface recombination velocity at higher temperature indicates that more carriers can reach the surface of the NWs and recombine there, which leads to weaker the hot carrier effects.

### III. CONCLUSION

In summary, we studied the dynamics of hot carriers in core-shell InGaAs/InAlAs NWs influenced by Auger heating and SRH recombination at various lattice temperatures and excitation power densities using steady-state and time-resolved PL spectroscopy. The results indicate that at 10 K, Auger heating in the NWs is dominant and it shows strong dependencies on the NW diameter, following a non-monotonic behavior from thick to thin NWs. Such strong Auger heating can enhance hot carrier temperature and improve the efficiency of hot carrier solar cells fabricated from such core-shell NWs. However, by increasing the lattice temperature, the contribution of SRH recombination becomes higher, resulting in increased rates of hot carrier relaxation. In addition, our results indicate that by raising the lattice temperature, the surface recombination velocity of carriers increases, which can further enhance the thermalization rates of hot carriers. As such, reduced hot carrier effects are observed at elevated lattice temperatures, and to mitigate this it will be important to optimize both the bulk and interface quality of the core-shell NWs towards minimized defect and impurity densities in the future.